\documentclass[12pt,titlepage]{article}
\usepackage{amsmath}
\usepackage{amssymb}
\usepackage{graphicx}
\usepackage{subfigure}

\newcommand{\bbf}[1]{\mbox{\boldmath $#1$}}
\newcommand{\indep}{{\bot\negthickspace\negthickspace\bot}}

\begin{document}
\title{Mediation Analysis Without Sequential Ignorability: Using Baseline
  Covariates Interacted with Random Assignment as Instrumental Variables}
\thanks{Dylan S. Small is Associate Professor, Department of
  Statistics, The Wharton School of the University of Pennsylvania,
  Philadelphia, PA 19104 (E-mail:
  {\it{dsmall@wharton.upenn.edu}}).}
\author{Dylan S. Small\\University of Pennsylvania}
\date{\today}
\maketitle

\begin{abstract}
In randomized trials, researchers are often interested in mediation analysis to understand how a treatment works, in particular how much of a treatment's effect is mediated by an intermediated variable and how much the treatment directly affects the outcome not through the mediator.  The standard regression approach to mediation analysis assumes sequential ignorability of the mediator, that is that the mediator is effectively randomly assigned given baseline covariates and the randomized treatment.  Since the experiment does not randomize the mediator, sequential ignorability is often not plausible.  Ten Have et al. (2007, {\it{Biometrics}}), Dunn and Bentall (2007, {\it{Statistics in Medicine}}) and Albert (2008, {\it{Statistics in Medicine}}) presented methods that use baseline covariates interacted with random assignment as instrumental variables, and do not require sequential ignorability.  We make two contributions to this approach.  First, in previous work on the instrumental variable approach, it has been assumed that the direct effect of treatment and the effect of the mediator are constant across subjects; we allow for variation in effects across subjects and show what assumptions are needed to obtain consistent estimates for this setting.  Second, we develop a method of sensitivity analysis for violations of the key assumption that the direct effect of the treatment and the effect of the mediator do not depend on the baseline covariates.
\vspace{1cm}

Keywords: Causal Inference, Mediation Analysis, Instrumental Variables.
\end{abstract}

\bf 1. Introduction \rm

Randomized trials are explicitly designed to estimate the effects of
treatments but not how those effects occur.  Yet, many
researchers are interested in how treatments that are evaluated using
randomized experiments achieve their effects.  Mediation analysis seeks to open up the ``black box'' of a treatment
and explain how it works. For example, the
PROSPECT study (Bruce {\it et al}., 2004) evaluated an intervention
for improving treatment of depression in the elderly in primary care
practices.  The intervention consisted of having a depression
specialist (typically a master's-level clinician) closely collaborate
with the depressed patient and the patient's primary care physician to
facilitate patient and clinician adherence to a treatment algorithm
and provide education, support and ongoing assessment to the patient.  The intervention significantly reduced depression (as measured by the Hamilton test) four months after baseline.
Researchers of this study are interested in to what extent the effect
of the intervention can be explained by its increasing use of
prescriptive anti-depressant medication as compared to other factors.
Understanding the mechanism by which a treatment achieves its effects
can help researchers and policymakers design more effective
treatments (Gennetian, Bos and Morris, 2002; Kraemer et al.,
2002).  For example, if the PROSPECT study intervention achieves its effects primarily through increasing use of antidepressants,
then a more cost-effective intervention might be designed that has the
depression specialist focus her time only on increasing use of antidepressants.

The standard approach to mediation analysis (Judd and Kenny, 1981; Baron and Kenny, 1986; MacKinnon et al.,
2002) makes a strong {\it{sequential ignorability}} assumption that, in addition to the intervention being randomly assigned, the mediating variable (e.g., antidepressant use) is also effectively randomly assigned given the assigned intervention and the measured confounding variables (i.e., the mediating variable is sequentially ignorable, meaning that there are no unmeasured confounders of the mediating variable-outcome relationship) (Ten Have et al., 2007).  In the PROSPECT study, potential unmeasured confounders of the mediating variable (antidepressant use)-outcome (depression) relationship include medical comorbidities during the follow-up period, which deter elderly depressed patients from taking antidepressant medications because of so many other medications that are necessitated by their medical comorbidities and also predisposes patients to more depression (Ten Have et al., 2007).  To address such unmeasured confounding, Ten Have et al. (2007) develop an alternative approach
to mediation analysis that allows for unmeasured confounding but relies on having a baseline covariate that
interacts with random assignment in predicting the mediating variable, and does not modify the effects of the mediating variable or the direct effect of the randomized treatment.
For example, for the PROSPECT study, Ten Have et al. considered the following baseline
covariates: baseline depression and baseline suicide ideation.  Ten Have et al.'s approach to mediation analysis uses a
rank preserving model for causal effects and g-estimation (Robins, 1994).  The assumption underlying Ten Have et al.'s approach, that there is a baseline covariate that interacts with random assignment in predicting the mediating variable but that does not modify the effect of the mediating variable or the direct effect of the randomized assigned treatment, can be viewed as an assumption that the baseline covariate interacted with random assignment is an instrumental variable (IV) for the mediating variable in a structural equation model.  Dunn and Bentall (2007) show that two stage least squares estimation of this structural equation model with the baseline covariate interacted with random assignment as an IV produces essentially equivalent results to that of $g$-estimation of the rank preserving model.  Gennetian, Bos and Morris (2002), Albert (2008) and Joffe et al. (2008) provide further discussion of this two stage least squares approach.

This paper makes two contributions to the approach of using baseline covariates interacted with random assignment as IVs for mediation analysis when sequential ignorability does not hold.  First, in previous work on the instrumental variable approach, it has been assumed that the effect of the mediator and the direct effect of treatment are constant across subjects; we allow for variation in effects across subjects and show what assumptions are needed to obtain consistent estimates for this setting.  Second, we develop a method of sensitivity analysis for violations of the key assumption that the direct effect of the treatment and the effect of the mediator do not depend on the baseline covariates.

Our paper is organized as follows.  Section 2 provides the notation and setup.  Section 3 describes the model we will consider.  Section 4 reviews the standard regression approach to mediation analysis.  Section 5 presents the instrumental variables approach.   Section 6 develops a method of sensitivity analysis for the effect of departures from the key assumption that the baseline covariate does not modify the causal effects of the random assignment or the mediating variable.  The methods are applied to the PROSPECT study.

\bf 2. Setup and Notation \rm

We assume there are $N$ subjects who are an iid sample from a population.  We assume that the treatment $R$ is randomized.

The observed variables for subject $i$ are the following: $Y_i$ is the observed outcome, $R_i$ is the observed randomized zero-one treatment assignment, ${\bf{X}}_i$ is a vector of observed baseline covariates other than treatment assignment and $M_i$ is the observed mediation variable.  The potential outcomes for subject $i$ are $Y_i^{(r,m)}$, $r=0$ or $1$ and $m\in {\mathcal{M}}$ where ${\mathcal{M}}$ is the set of possible values the mediating variable can take on; $Y_i^{(r,m)}$ is the outcome variable that would be observed if subject $i$ were randomized to level $r$ of the treatment and through some hypothetical mechanism were to receive or exhibit level $m$ of the mediator.  To establish a unique potential outcome, we assume that all such hypothetical mechanisms lead to the same potential outcome (Ten Have et al., 2007).  The observed outcome $Y_i$ is equal to $Y_i^{(R_i,M_i)}$. The potential mediating variables for subject $i$ are $M_i^{(r)}$, $r=0$ or $1$; $M_i^{(r)}$ is the level of the level of the mediating variable that would be observed if subject $i$ were assigned level $r$ of the treatment.  The observed mediating variable $M_i$ equals $M_i^{(R_i)}$.

We let the random variables $Y, R, {\bf{X}}, Y^{(r,m)} (r=0,1, m\in {\mathcal{M}}), M^{(r)} (r=0,1)$ be the values of the observed outcome, treatment assignment, baseline covariates, potential outcomes and potential mediating variables for a randomly chosen subject from the population.

\bf 3. Model \rm

We consider the following model for potential outcomes:
\begin{equation}
Y_i^{(r,m)}=Y_i^{(0,0)}+\theta_{M_i} m+\theta_{R_i} r, \label{potential.outcome.model.1}
\end{equation}
where the $(Y_i^{(0,0)},\theta_{M_i},\theta_{R_i})$ are iid random vectors.  Here $\theta_{M_i}$ represents the effect for subject $i$ of a one unit increase in the mediator on the outcome holding the treatment fixed at any level $r$.  The parameter $\theta_{R_i}$ represents the direct effect for subject $i$ of the treatment on the outcome holding the mediator fixed at any level $m$.  Let $\theta_M=E(\theta_{M_i})$ be the average effect of a one unit increase in the mediator and $\theta_R=E(\theta_{R_i})$ be the average direct effect of the treatment.


\bf 4. Review of Standard Regression Approach\rm

The standard regression approach of Baron and Kenny (1986) is to estimate $\theta_M$ and $\theta_R$ by least squares regression of $Y_i$ on $M_i$ and $R_i$. Under the maintained assumption that $R$ is randomized, the standard regression approach provides consistent estimates of $\theta_M$ and $\theta_R$ under the additional assumption that $M$ is sequentially ignorable given $R$:
\begin{equation}
M_i\indep Y_i^{(R_i,m)}, m\in {\mathcal{M}}, \label{path.analysis.assumption}
\end{equation}
where ${\mathcal{M}}$ is the set of possible values of the mediating variable $M$.  The sequentially ignorable assumption (\ref{path.analysis.assumption}) means that $M$ is effectively randomly assigned given $R$.  Under model (\ref{potential.outcome.model.1}), the sequential ignorability assumption (\ref{path.analysis.assumption}) is equivalent to
\begin{equation}
M_i\indep Y_i^{(0,0)},\theta_{M_i},\theta_{R_i}. \label{path.analysis.assumption.alternative}
\end{equation}
See Imai, Keele and Yamamoto (2010) for further discussion of the sequential ignorability assumption.  The sequentially ignorable assumption (\ref{path.analysis.assumption}) will be violated if there are confounders of the mediator-outcome relationship.  Measured baseline confounders of the mediator-outcome relationship can be controlled for by controlling for these confounders in the regression.  If there are measured postbaseline confounders, the regression on the measured confounders will produce an unbiased estimate of $\theta_M$ but not $\theta_R$; to obtain an unbiased estimate of $\theta_R$, $Y-\hat{\theta}_M$ can be regressed on $R$ (Vansteelandt, 2009; Ten Have and Joffe, 2010).

\bf 5. Instrumental Variables Approach \rm

The standard regression approach can only control for measured confounders of the mediator-outcome relationship.  The IV approach using baseline covariates interacted with treatment assignments can control for unmeasured confounders when baseline covariate(s) interacted with treatment assignment are valid IVs.  This IV approach for mediation analysis models has been discussed by Dunn and Bentall (2007) and Albert (2008), and the closely related $g$-estimation approach has been discussed by Ten Have et al. (2007).  These authors have considered models in which the direct effect of treatment and the effect of the mediating variable are the same for all subjects.  We will allow these effects to vary from subject to subject as in (\ref{potential.outcome.model.1}) and provide conditions needed for the instrumental variable to be consistent.

Denote a vector of baseline covariates by ${\bf{X}}$.  We assume that the association of ${\bf{X}}$ with the potential outcomes is linear:
\begin{equation}
E(Y^{(0,0)}|{\bf{X}})=\alpha + \bbf{\beta}^T{\bf{X}} \label{linear.model.y00}
\end{equation}
Then, we can write the observed data $Y_i$ as
\begin{gather}
Y_i = \bbf{\beta}^T{\bf{X}}_i+\theta_R R_i+\theta_M M_i + \epsilon_i, \nonumber \\
\epsilon_i = (\theta_{R_i}-\theta_R)R_i + (\theta_{M_i}-\theta_M)M_i + Y_i^{(0,0)}-E(Y_i^{(0,0)}|{\bf{X}}_i) \label{observed.data.iv.model} \end{gather}
The least squares regression of $Y$ on ${\bf{X}}$, $R$ and $M$ will produce biased estimates if there are unobserved confounders of the mediator-outcome relationship that make $\epsilon_i$ correlated with $M_i$.  The method of instrumental variables (IVs) seeks to replace $M_i$ with its expectation given instrumental variables that help to predict $M_i$ and are uncorrelated with $\epsilon_i$.  The interactions between the baseline covariates ${\bf{X}}$ and $R$ are valid IVs if the following conditions hold:
\begin{enumerate}
\item[(IV-A1)] The interaction between $R$ and ${\bf{X}}$ is helpful for predicting $M$ in a linear model, i.e., $E^*(M|R,{\bf{X}})\neq E^*(M|R,{\bf{X}},R{\bf{X}})$ where $E^*(M|{\bf{A}})=\arg\min_{\bbf{\lambda}} E(M-\bbf{\lambda}^T{\bf{A}})^2$ denotes the best linear predictor of $M$ given ${\bf{A}}$.
\item[(IV-A2)] The average direct effect of the treatment given ${\bf{X}}$, $E(\theta_{R_i}|{\bf{X}}_i={\bf{X}}$, is the same for all ${\bf{X}}$, i.e., $E(\theta_{R_i}|{\bf{X}}_i)={\bf{X}})=\theta_R$ for all ${\bf{X}}$.  Likewise, the average effect of the mediating variable given ${\bf{X}}$, $E(\theta_{M_i}|{\bf{X}}_i={\bf{X}})$, is the same for all ${\bf{X}}$, i.e., $E(\theta_{M_i}|{\bf{X}}_i={\bf{X}})=\theta_M$ for all ${\bf{X}}$.
\item[(IV-A3)] The value of the mediating variable is independent of the effect of the mediating variable given the treatment and the baseline covariates
    \begin{equation}
    M_i\indep \theta_{M_i}|R_i,{\bf{X}}_i \label{mediating.independent.effect}
    \end{equation}
\end{enumerate}
(IV-A1) says that $R{\bf{X}}$ helps to predict $M$.  (IV-A2) and (IV-A3), and the assumption that $R$ is randomly assigned, together guarantee that $R{\bf{X}}$ is uncorrelated with $\epsilon_i$, which we show in the following.

{\bf{Proposition 1}}: Under (IV-A2) and (IV-A3) and the assumption that $R$ is randomly assigned, each component of $R\times {\bf{X}}_i$ is uncorrelated with $\epsilon_i$.

{\bf{Proof}}: Consider a component of $R\times {\bf{X}}_i$, $RX_{i1}$. From (\ref{observed.data.iv.model}), $\epsilon_i=(\theta_{R_i}-\theta_R)R_i+(\theta_{M_i}-\theta_M)M_i+\{Y_i^{(0,0)}-E(Y_i^{(0,0)}|{\bf{X}}_i)\}$. We will prove that $Cov(RX_{i1},\epsilon_i)=0$ by showing that $RX_{i1}$ is uncorrelated with each of the three summands that make up $\epsilon_i$, namely (i) $Cov(RX_{i1},(\theta_{R_i}-\theta_R)R_i)=0$; (ii) $Cov(RX_{i1},(\theta_{M_i}-\theta_M)M_i)=0$ and (iii) $Cov(RX_{i1},Y_i^{(0,0)}-E(Y_i^{(0,0)}|{\bf{X}}_i))=0$.  For (i), since $R_i$ is randomized, we have $E[(\theta_{R_i}-\theta_R)R_i]=0$ so that $Cov(R_iX_{i1},(\theta_{R_i}-\theta_R)R_i)=E(R_iX_{i1}(\theta_{R_i}-\theta_R)R_i)$.  Furthermore, we have
\begin{eqnarray*}
E(R_iX_{i1}(\theta_{R_i}-\theta_R)R_i) & = & E(R_i^2)E(X_{i1}(\theta_{R_i}-\theta_R)) \\
& = & 0,
\end{eqnarray*}
where the first equality follows from the fact that $R$ is randomized and the second equality follows from (IV-A2).  This proves (i).  For (ii), we first note that
\begin{eqnarray*}
E[(\theta_{M_i}-\theta_M)M_i] & = & E[E[(\theta_{M_i}-\theta_M)M_i|R_i,{\bf{X}}_i]] \\
& = & E[E[(\theta_{M_i}-\theta_M)|R_i,{\bf{X}}_i]E[M_i|R_i,{\bf{X}}_i]] \\
& = & 0,
\end{eqnarray*}
where the second equality follows from (IV-A3) and the third equality follows from (IV-A2) and the fact that $R$ is randomized.  Thus,
$Cov(R_iX_{i1},(\theta_{M_i}-\theta_M)M_i)=E(R_iX_{i1}(\theta_{M_i}-\theta_M)M_i)$, and
\begin{eqnarray*}
E(R_iX_{i1}(\theta_{M_i}-\theta_M)M_i) & = & E[E[R_iX_{i1}(\theta_{M_i}-\theta_M)M_i|R_i,{\bf{X}}_i]] \\
& = & E[R_iX_{i1}E[(\theta_{M_i}-\theta_M)M_i|R_i,{\bf{X}}_i] \\
& = & E[R_iX_{i1}E[(\theta_{M_i}-\theta_M)|R_i,{\bf{X}}_i]E[M_i|R_i,{\bf{X}}_i]] \\
& = & 0,
\end{eqnarray*}
where the third equality follows from (IV-A3) and the fourth equality follows from (IV-A2) and the fact that $R$ is randomized.  This proves (ii).  For (iii),
\begin{eqnarray*}
Cov(R_iX_{i1},Y_i^{(0,0)}-E[Y_i^{(0,0)}|{\bf{X}}_i]) & = & E[R_iX_{i1}\{Y_i^{(0,0)}-E[Y_i^{(0,0)}|{\bf{X}}_i]\}] \\
& = & E(R_i)E[X_{i1}\{Y_i^{(0,0)}-E[Y_i^{(0,0)}|{\bf{X}}_i]\}] \\
& = & 0,
\end{eqnarray*}
where the second equality follows from $R$ being randomized and third equality from properties of conditional expectation.  This proves (iii).  $\Box$

Assumption (IV-A3) is weaker than the sequential ignorability assumption (\ref{path.analysis.assumption}) because (IV-A3) does not say that $Y_i^{(0,0)}$ is independent of $M_i$.  Assumption (IV-A3) says that the level of the mediating variable is independent of the effect the mediating variable has, while sequential ignorability says that not only is the level independent of the effect, but also the level is independent of all the person's potential outcomes.  In the context of the PROSPECT study, (IV-A3) says that antidepressant use is independent of the effect that the antidepressant would have, while sequential ignorability says that not only is antidepressant use independent of its effect, but antidepressant use is also independent of unmeasured medical comorbidities and any other unmeasured variables that affect depression.  Note that (IV-A3) is automatically satisfied if $\theta_{R_i}$ and $\theta_{M_i}$ if $\theta_{R_i}$ and $\theta_{M_i}$ are the same for all subjects as is assumed by Ten Have et al. (2007), Dunn and Bentall (2007) and Albert (2008).

Under (IV-A2)-(IV-A3), we have
\begin{eqnarray*}
E^*(Y|R,{\bf{X}},R\times {\bf{X}}) & = & \alpha + \bbf{\beta}^T{\bf{X}}+\theta_{R}R+\theta_{M}E^*(M|R,{\bf{X}},R\times {\bf{X}})+E^*(\epsilon|R,{\bf{X}},R\times{\bf{X}})\\
& = & \alpha + \bbf{\beta}^T{\bf{X}}+\theta_{R}R+\theta_{M}E^*(M|R,{\bf{X}},R\times {\bf{X}}),
\end{eqnarray*}
The two-stage least squares estimates of $\theta_R$ and $\theta_M$ are found as follows:
\begin{enumerate}
\item Regress $M$ on $R$, ${\bf{X}}$ and $R\times {\bf{X}}$ using least squares and obtain the predicted values $\hat{E}(M|R,{\bf{X}},R\times {\bf{X}})$.
\item Regress $Y$ on $R$, ${\bf{X}}$ and $\hat{E}(M|R,{\bf{X}},R\times {\bf{X}})$ using least squares.  The coefficient on $R$ is $\hat{\theta}_R$ and the coefficient on $\hat{E}(M|R,{\bf{X}},R\times {\bf{X}})$ is $\hat{\theta}_M$.
    \end{enumerate}
Using the theory of instrumental variables for single-equation linear models (Wooldridge, 2002, Ch. 5),
the two stage least squares estimates are consistent under (IV-A1)-(IV-A3) because (i) $\mbox{Cov}(R\times {\bf{X}},\epsilon)={\bf{0}}$ under (IV-A2)-(IV-A3) and (ii) the coefficient on $R\times{\bf{X}}$ in the linear projection of $Y$ onto $R$, ${\bf{X}}$ and $R\times{\bf{X}}$ is not ${\bf{0}}$ under (IV-A1).

We now discuss the variance-covariance matrix of $\hat{\bbf{\kappa}}=(\hat{\alpha},\hat{\bbf{\beta}},\hat{\theta}_R,\hat{\theta}_M)$.  First, consider the following additional assumptions:
\begin{enumerate}
\item[(AA-1)] The distribution of the direct effect of the treatment and the effect of the mediating variable do not depend on ${\bf{X}}_i$,
    \[
\theta_{R,i},\theta_{M,i}\indep {\bf{X}}_i.
\]
\item[(AA-2)] $\mbox{Var}(\{ Y_i^{(0,0)}-E(Y_i^{(0,0)})\}|{\bf{X}}_i={\bf{X}})$ is the same for all ${\bf{X}}$.
    \end{enumerate}
Under (AA-1)-(AA-2), the $Var(\epsilon_i|R_i,{\bf{X}}_i)$ is the same for all $R_i, {\bf{X}}_i$.  Then a consistent estimate of the variance-covariance matrix of $\hat{\bbf{\kappa}}$ is
$\hat{\sigma}_{\epsilon}^2 ({\bf{A}}^T{\bf{A}})^{-1}$ where $\hat{\sigma}_{\epsilon}^2=\frac{1}{N}\sum_{i=1}^N \hat{\epsilon}_i^2$, $\hat{\epsilon}_i=Y_i-\hat{\alpha}-\hat{\bbf{\beta}}^T{\bf{X}}_i-\hat{\theta}_R R_i-\hat{\theta}_MM_i$ and ${\bf{A}}$ is a matrix with $N$ rows consisting of a column of ones, columns for each of the variables in ${\bf{X}}$ for the $N$ subjects, a column of the values of $R$ for the $N$ subjects and a column of the values of $\hat{E}^*(M|R,{\bf{X}},R\times {\bf{X}})$ for the $N$ subjects (Wooldridge, 2002, Ch. 5).  By a consistent estimate of the covariance matrix, we mean that $\sqrt{N}\hat{Cov}(\hat{\bbf{\kappa}}_N)$ is a consistent estimator of $\sqrt{N}Cov(\hat{\bbf{\kappa}}_N)$, where $\hat{\bbf{\kappa}}_N$ is the two stage least squares estimator of $\bbf{\kappa}$ based on $N$ observations.

Suppose that either (a) the $Y_i^{(0,0)}-E(Y_i^{(0,0)}|{\bf{X}}_i)$ have a distribution that depends on ${\bf{X}}_i$; and/or (b) the direct effects of treatment and the effect of the mediating variable have a distribution that might depend on ${\bf{X}}$ but the mean is the same for all ${\bf{X}}$, i.e., $E(\theta_{R,i}|{\bf{X}}_i)=\theta_R$ and $E(\theta_{M,i}|{\bf{X}}_i)=\theta_M$. Then, the two stage least squares estimate remains consistent, but the usual standard error might be inconsistent.  A consistent estimate of the covariance matrix under regularity conditions (White, 1982; Wooldridge, 2002, Ch. 5.2.5) is the ``sandwich'' estimator, $({\bf{A}}^T{\bf{A}})^{-1}\left (\sum_{i=1}^N \hat{\epsilon}_i^2{\bf{A}}_i^T{\bf{A}}_i\right)({\bf{A}}^T{\bf{A}})^{-1}$, where ${\bf{A}}_i=(1,{\bf{X}}_i,R_i,M_i)^T$.

Inferences from two stage least squares become unreliable if the IV(s) are ``weak,'' which in our setting means that the interaction between $R$ and ${\bf{X}}$ is only a weak predictor of $M$ in the linear model, i.e., $E^*(M|R,{\bf{X}},R{\bf{X}})$. Specifically, when the IV(s) are weak, the two stage least squares estimates can have a large bias in the direction of the ordinary least squares estimates of $Y$ on ${\bf{X}}$, $R$ and $M$, and the coverage of the confidence intervals for the two stage least squares estimates can be poor (Bound, Jaeger and Baker, 1995).  Stock, Wright and Yogo (2002) provided a criterion for when IV inference is reliable based on the partial $F$ statistic for testing that the coefficient on the $R\times {\bf{X}}$ variable are zero from the first stage regression of $M$ on $R$, ${\bf{X}}$ and $R\times {\bf{X}}$.  Inference can be expected to reliable when this $F$ statistic is greater than 8.96, 11.59, 12.83, 15.09, 20.88 and 26.80 for 1, 2, 3, 5, 10 and 15 variables in ${\bf{X}}$ respectively.  This
criterion is based on the goal of having a nominal $0.05$ level test of the coefficient on $M$ have at most actual level $0.15$, and the chance that we falsely say that a nominal $0.05$ level test of $M$ has at most actual level $0.15$ be at most $0.05$.

In our notation, we have assumed that all of the baseline variables ${\bf{X}}$ that we control for are interacted with the randomized intervention $R$ to form instrumental variables.  We might want to control for additional baseline variables ${\bf{Z}}$ that we do not think satisfy (IV-A2); controlling for these additional baseline variables might increase precision.  In order to control for such additional baseline variables ${\bf{Z}}$, we include ${\bf{Z}}$ in both the first and second stage regressions but do not use $R\times {\bf{Z}}$ as instrumental variables.

\bf 5.1 Application to PROSPECT study \rm

We use the PROSPECT study data set provided by Ten Have et al. (2007) under the Article Information link at the {\it{Biometrics}} website http://www.tibs.org/biometrics.  There are 297 subjects, 145 were randomized to the intervention and 152 to the control.  The outcome is the subject's Hamilton score (a measure of depression, with a higher score indicating more depression) four months after the intervention.  Figure 1 shows the distribution of the outcome in the intervention and control groups.

\begin{figure}
\centerline{\includegraphics[width=5in,height=5in]{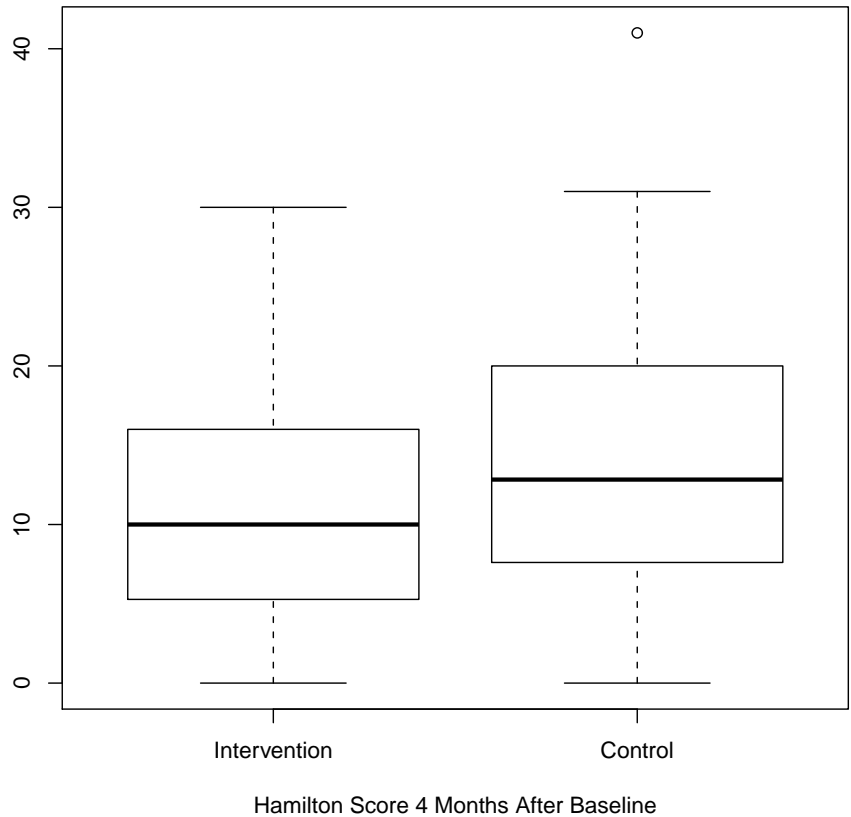}}
\caption{Box plots of the outcome in the intervention and control groups.}
\end{figure}

The mediating variable is an indicator for whether the subject used antidepressants during the period from the intervention to four months after the intervention.  The intervention significantly increases the mediator -- the intervention is estimated to multiply the odds of antidepressant use by $6.7$ with a 95\% confidence interval of (3.9, 11.7).

The second row of Table 1 shows estimates from the standard regression approach.  The baseline covariates used are (i) an indicator of whether the subject had used antidepressants in the past and (ii) a baseline ordinal measure of antidepressant use that ranges from 0 (no baseline use of antidepressants) to 4 (highest level of baseline use of antidepresants).  The intervention is estimated to have a direct effect of reducing depression and antidepressant use is estimated to reduce depression, but neither effect is significant.

Following Ten Have et al. (2007), we consider as instrumental variables the interaction between the randomized intervention and the baseline covariates.  The partial $F$ statistic for the instruments in the first stage regression is $27.13$ indicating that these are not weak instruments.   The two stage least squares estimates are shown in the third row of Table 2.
The confidence intervals are based on the assumption that the $\epsilon_i$ are homoskedastic, but the confidence intervals are similar if we use the sandwich covariance estimates that allow for heteroskedasticity.
\begin{table}[h!]
\begin{center}
\begin{tabular}{|c|c|c|}
\hline Method & Direct effect of intervention & Mediator effect \\
Standard Regression & -1.67 (-3.69, 0.36) & -1.02 (-3.40, 1.36) \\
IV & -0.94 (-3.92,2.04) & -2.87 (-8.89, 3.15) \\
\hline
\end{tabular}
\caption{Estimates for the direct effect of the intervention and the mediator (antidepressant use) effect in the PROSPECT study.  95\% confidence intervals are in parentheses.}
\end{center}
\end{table}

\bf 6. Sensitivity Analysis \rm

In this section, we will consider the sensitivity of inferences to violations of assumption (IV-A2) that the average direct effect of the treatment given ${\bf{X}}$ and the average effect of the mediating variable given ${\bf{X}}$ are the same for all ${\bf{X}}$.  Consider the following parametric family of violations of assumption (IV-A2):
\begin{gather}
E[\theta_{R_i}|{\bf{X}}_i={\bf{X}}]=\theta_R+\bbf{\tau}_R^T({\bf{X}}_i-E[{\bf{X}}]), \nonumber \\
E[\theta_{M_i}|{\bf{X}}_i={\bf{X}}]=\theta_M+\bbf{\tau}_M^T({\bf{X}}_i-E[{\bf{X}}]). \label{sensitivity.model}
\end{gather}

(IV-A2) is satisified if $\bbf{\tau}_R={\bf{0}}$ and $\bbf{\tau}_M={\bf{0}}$.  Suppose we know the value of $\bbf{\tau}_R$, $\bbf{\tau}_M$ and $E[{\bf{X}}]$.  Then, we can write,
\begin{gather}
Y_i -R_i\bbf{\tau}_R^T({\bf{X}}_i-E[{\bf{X}}])-M_i\bbf{\tau}_M^T({\bf{X}}-E[{\bf{X}}])= \bbf{\beta}^T{\bf{X}}_i+\theta_R R_i+\theta_M M_i + \epsilon_i, \nonumber \\
\epsilon_i = (\theta_{R_i}-E(\theta_{R_i}|{\bf{X}}_i))R_i + (\theta_{M_i}-E(\theta_{M_i}|{\bf{X}}_i))M_i + Y_i^{(0,0)}-E(Y_i^{(0,0)}|{\bf{X}}_i) \label{sensitivity.iv.model}
\end{gather}

Now, we show that $R_i\times {\bf{X}}_i$ are valid IVs for estimating $\theta_R$ and $\theta_M$ when the response variable is $Y_i -\bbf{\tau}_R^T({\bf{X}}_i-E[{\bf{X}}])-\bbf{\tau}_M^T({\bf{X}}_i-E[{\bf{X}}]$.

{\bf{Proposition 2}}: Under (\ref{sensitivity.model}), (IV-A3) and the assumption that $R$ is randomly assigned, each component of $R\times {\bf{X}}_i$ is uncorrelated with $\epsilon_i$.

{\bf{Proof}}: Consider a component of $R\times {\bf{X}}_i$, $RX_{i1}$. From (\ref{sensitivity.iv.model}), $\epsilon_i=(\theta_{R_i}-E(\theta_{R_i}|{\bf{X}}_i))R_i+(\theta_{M_i}-E(\theta_{M_i}|{\bf{X}}_i))M_i+\{Y_i^{(0,0)}-E(Y_i^{(0,0)}|{\bf{X}}_i)\}$. We will prove that $Cov(RX_{i1},\epsilon_i)=0$ by showing that $RX_{i1}$ is uncorrelated with each of the three summands that make up $\epsilon_i$, namely (i) $Cov(RX_{i1},(\theta_{R_i}-E(\theta_{R_i}|{\bf{X}}_i))R_i)=0$; (ii) $Cov(RX_{i1},(\theta_{M_i}-E(\theta_{M_i}|{\bf{X}}_i))M_i)=0$ and (iii) $Cov(RX_{i1},Y_i^{(0,0)}-E(Y_i^{(0,0)}|{\bf{X}}_i))=0$.  For (i), since $R_i$ is randomized, we have $E[(\theta_{R_i}-E(\theta_{R_i}|{\bf{X}}_i))R_i]=0$ so that $Cov(R_iX_{i1},(\theta_{R_i}-E(\theta_{R_i}|{\bf{X}}_i))R_i)=E(R_iX_{i1}(\theta_{R_i}-E(\theta_{R_i}|{\bf{X}}_i))R_i)$.  Furthermore, we have
\begin{eqnarray*}
E(R_iX_{i1}(\theta_{R_i}-E(\theta_{R_i}|{\bf{X}}_i))R_i) & = & E(R_i^2)E(X_{i1}(\theta_{R_i}-E(\theta_{R_i}|{\bf{X}}_i))) \\
& = & 0,
\end{eqnarray*}
where the first equality follows from the fact that $R$ is randomized and the second equality follows from properties of conditional expectation.  This proves (i).  For (ii), we first note that
\begin{eqnarray*}
E[(\theta_{M_i}-E(\theta_{M_i}|{\bf{X}}_i))M_i] & = & E[E[(\theta_{M_i}-E(\theta_{M_i}|{\bf{X}}_i))M_i|R_i,{\bf{X}}_i]] \\
& = & E[E[\theta_{M_i}-E(\theta_{M_i}|{\bf{X}}_i)|R_i,{\bf{X}}_i]E[M_i|R_i,{\bf{X}}_i]] \\
& = & 0,
\end{eqnarray*}
where the second equality follows from (IV-A3) and the third equality follows from the fact that $R$ is randomized and properties of conditional expectation.  Thus,
$Cov(R_iX_{i1},(\theta_{M_i}-E(\theta_{M_i}|{\bf{X}}_i))M_i)=E(R_iX_{i1}(\theta_{M_i}-E(\theta_{M_i}|{\bf{X}}_i))M_i)$, and
\begin{eqnarray*}
E(R_iX_{i1}(\theta_{M_i}-E(\theta_{M_i}|{\bf{X}}_i))M_i) & = & E[E[R_iX_{i1}(\theta_{M_i}-E(\theta_{M_i}|{\bf{X}}_i))M_i|R_i,{\bf{X}}_i]] \\
& = & E[R_iX_{i1}E[(\theta_{M_i}-E(\theta_{M_i}|{\bf{X}}_i))M_i|R_i,{\bf{X}}_i]] \\
& = & E[R_iX_{i1}E[(\theta_{M_i}-E(\theta_{M_i}|{\bf{X}}_i))|R_i,{\bf{X}}_i]E[M_i|R_i,{\bf{X}}_i]] \\
& = & 0,
\end{eqnarray*}
where the third equality follows from (IV-A3) and the fourth equality follows from the fact that $R$ is randomized and properties of conditional expectation.  This proves (ii).  For (iii),
\begin{eqnarray*}
Cov(R_iX_{i1},Y_i^{(0,0)}-E[Y_i^{(0,0)}|{\bf{X}}_i]) & = & E[R_iX_{i1}\{Y_i^{(0,0)}-E[Y_i^{(0,0)}|{\bf{X}}_i]\}] \\
& = & E(R_i)E[X_{i1}\{Y_i^{(0,0)}-E[Y_i^{(0,0)}|{\bf{X}}_i]\}] \\
& = & 0,
\end{eqnarray*}
where the second equality follows from $R$ being randomized.  This proves (iii).  $\Box$

Based on Proposition 2, we can make inferences for $\theta_R$ and $\theta_M$ under (IV-A1), (IV-A3) and (\ref{sensitivity.model}) by replacing $Y_i$ by $Y_i -R_i\bbf{\tau}_R^T({\bf{X}}_i-E[{\bf{X}}])-M_i\bbf{\tau}_M^T({\bf{X}}-E[{\bf{X}}])$ in the two stage least squares inference procedure from Section 5.  Specifically, for given values of $\bbf{\tau}_R$ and $\bbf{\tau}_M$, we regress $Y -R\bbf{\tau}_R^T({\bf{X}}-E[{\bf{X}}])-M\bbf{\tau}_M^T({\bf{X}}-E[{\bf{X}}])$ on $R$, ${\bf{X}}$ and $\hat{E}(M|R,{\bf{X}},R\times {\bf{X}})$ using least squares.  Then, the estimated values of $\theta_R$ and $\theta_M$ given $\bbf{\tau}_R,\bbf{\tau}_M$ are the coefficients on $R$ and
$\hat{E}(M|R,{\bf{X}},R\times {\bf{X}})$ respectively.  The variance-covariance matrix of the estimate of $\bbf{\kappa}=(\alpha ,\bbf{\beta},\theta_R,\theta_M)$ given $\theta_R$ and $\theta_M$ is $\hat{\sigma}_{\epsilon}^2 ({\bf{A}}^T{\bf{A}})^{-1}$ where now $\hat{\sigma}_{\epsilon}^2=\frac{1}{N}\sum_{i=1}^N [Y_i -R_i\bbf{\tau}_R^T({\bf{X}}_i-E[{\bf{X}}])-M_i\bbf{\tau}_M^T({\bf{X}}-E[{\bf{X}}])-\hat{\alpha}-\bbf{\beta}^T{\bf{X}}_i-\hat{\theta}_R R_i-\hat{\theta}_M M_i]^2$.

To carry out a sensitivity analysis for possible violations of the assumption (IV-A2) that the average direct effect of the treatment given ${\bf{X}}$ and the average effect of the mediating variable given ${\bf{X}}$ are the same for all ${\bf{X}}$, we consider how inferences vary over plausible values of ${\bbf{\tau}}_R$ and ${\bbf{\tau}}_M$.  The sensitivity parameters $\bbf{\tau}_R$ and $\bbf{\tau}_M$ have the following interpretation: the $j$th component of $\bbf{\tau}_R$ says how much does a one unit increase in the $j$th component of ${\bf{X}}$ change the direct effect of the treatment; the $j$th component of $\bbf{\tau}_M$ says how much does a one unit increase in the $j$th component of ${\bf{X}}$ change the effect of the mediator.  Shepherd, Gilbert and Mehrotra (2007) discuss methods for eliciting plausible values of sensitivity parameters from subject matter experts. 

Table 2 shows the results of a sensitivity analysis for the PROSPECT study.  We considered values of $\bbf{\tau}_R$ that allowed for the direct effect of the treatment to increase by one point for subjects who used antidepressants in the past compared to those subjects who did not use antidepressants in the past and the direct effect of the treatment to increase by one point for subjects who had a one category higher baseline use of antidepressants; we also considered values of $\bbf{\tau}_M$ that allowed for the effect of the mediator to be one point higher for subjects who used antidepressants in the past compared to those subjects who did not use antidepressants in the past and the effect of the mediator to be one point higher for subjects who had a one category higher baseline use of antidepressants.  Table 2 shows that inferences about the direct effect of the intervention and the mediator effect are fairly sensitive to violations of the assumption (IV-A2) in the range considered.  The point estimates of the direct effect of the intervention range from -3.63 to 0.94 and the point estimates of the mediator effect range from -2.87 to 5.33.

\begin{table}[h!]
\begin{center}
\begin{tabular}{|c|c|c|c|}
\hline $\bbf{\tau}_{R}$ & $\bbf{\tau}_M$ & Direct effect of intervention & Mediator effect \\
\hline (0,0) & (0,0) & -0.94 (-3.92,2.04) & -2.87 (-8.89, 3.15) \\
(0,1) & (0,0) & -2.75 (-5.73, 0.24) & 1.73 (-4.03, 7.76) \\
(1,0) & (0,0) & -1.58 (-4.54, 1.39) & -1.24 (-7.24, 4.77) \\
(1,1) & (0,0) & -3.39 (-6.39, -0.38) & 3.36 (-2.73, 9.45) \\
(0,0) & (0,1) & -1.03 (-4.00, 1.94) & -1.62 (-7.63, 4.40) \\
(0,1) & (0,1) & -2.84 (-5.84, 0.16) & 2.98 (-3.09, 9.05) \\
(1,0) & (0,1) & -1.67 (-4.64, 1.30) & 0.02 (-5.99, 6.02) \\
(1,1) & (0,1) & -3.48 (-6.51, -0.45) & 4.61 (-1.52, 10.75) \\
(0,0) & (1,0) & -1.09 (-4.06, 1.88) & -2.16 (-8.18, 3.86) \\
(0,1) & (1,0) & -2.90 (-5.89, 0.09) & 2.44 (-3.61, 8.49) \\
(1,0) & (1,0) & -1.73 (-4.70, 1.24) & -0.52 (-6.53, 5.48) \\
(1,1) & (1,0) & -3.54 (-6.57, -0.51) & 4.07 (-2.05, 10.20) \\
(0,0) & (1,1) & -1.18 (-4.16, 1.79) & -0.91 (-6.92, 5.11) \\
(0,1) & (1,1) & -2.99 (-6.01, 0.02) & 3.69 (-2.41, 9.79) \\
(1,0) & (1,1) & -1.82 (-4.80, 1.15) & 0.73 (-5.29, 6.75) \\
(1,1) & (1,1) & -3.63 (-6.69, -0.58) & 5.33 (-0.86, 11.51) \\ 
\hline
\end{tabular}
\caption{Estimates for the direct effect of the intervention and the mediator (antidepressant use) effect in the PROSPECT study under different values of the sensitivity parameters $\bbf{\tau}_R$ and $\bbf{\tau}_M$.  The first component of $\bbf{\tau}_R$ and $\bbf{\tau}_M$ corresponds to past antidepressant use and the second component corresponds to baseline antidepressant use.  95\% confidence intervals are in parentheses.}
\end{center}
\end{table}

\bf 7. Discussion \rm

The standard regression approach to mediation analysis assumes sequential ignorability of the mediator, that is that the mediator is effectively randomly assigned given baseline covariates and the randomized treatment.  Since the experiment does not randomize the mediator, sequential ignorability is often not plausible.  Ten Have et al. (2007, {\it{Biometrics}}), Dunn and Bentall (2007, {\it{Statistics in Medicine}}) and Albert (2008, {\it{Statistics in Medicine}}) presented methods that use baseline covariates interacted with random assignment as instrumental variables, and do not require sequential ignorability.  In this paper, we have discussed the setting in which there is variation in effects across subjects and shown what assumptions are needed to obtain consistent estimates for this setting when using baseline covariates interacted with random assignment as instrumental variables.  We have also developed a method of sensitivity analysis for violations of the assumption that the baseline covariates interacted with random assignment are valid instrumental variables, in particular violations of the assumption that the direct effect of the treatment and the effect of the mediator do not depend on the baseline covariates.  Gennetian, Bos and Morris (2002) have discussed baseline covariates that might be approximately valid instrumental variables when interacted with the randomized intervention, such as site in a multisite randomized experiments and baseline characteristics such as age or gender.  These authors also identified potential concerns that the effect of the mediator or the direct effect of the treatment might vary with these baseline variables.  Our sensitivity analysis method is useful for quantifying what inferences can be made under plausible violations of the assumption that the effect of the mediator or the direct effect of the treatment does not vary with baseline characteristics.

\vspace{1cm}

\bf Dedication. \rm  This paper is dedicated to my friend and mentor Tom Ten Have.  Tom provided a lot of insightful suggestions in the early stage of this work, and unfortunately passed away before I could discuss the later stages of the work with him.
\bf \center{References} \rm
\begin{description}\addtolength{\itemsep}{-0.5\baselineskip}
\item Albert, J.M. (2008).  Mediation analysis via potential outcomes models.  {\it{Statistics in Medicine}}, 27, 1282-1304.
\item Baron, R.M. and Kenny, D.A. (1986).  The moderator-mediator variable distinction in social psychological research: conceptual, strategic and statistical considerations.  {\it{Journal of Personality and Social Psychology}}, 51, 1173-1182.
\item Bound, J., Jaeger, D.A. and Baker, R.M. (1995).  Problems with Instrumental Variables Estimation When the Correlation Between the Instruments and the Endogeneous Explanatory Variable is Weak.  {\it{Journal of the American Statistical Association}}, 90, 443-450.
\item Bruce, M., Ten Have, T.R., Reynolds, C. et al. (2004).  A randomized trial to reduce suicidal ideation and depressive symptoms in depressed older primary care patients: The PROSPECT study.  {\it{Journal of the American Medical Association}}, 291, 1081-1091.
\item Dunn, G. and Bentall, R. (2007).  Modeling treatment effect heterogeneity in randomised controlled trials of complex interventions (psychological treatments).  {\it{Statistics in Medicine}}, 26, 4719-4745.
\item Gennetian, L., Bos, J. and Morris, P. (2002).  Using instrumental variables to learn more from social policy experiments.  MDRC Working Papers on Research Methodology.
\item Joffe, M., Small, D., Brunelli, S., Ten Have, T. and Feldman, H. (2008). Extended instrumental variables estimation for overall effects . {\it{International Journal of Biostatistics}}, 4(1), Article 4.
\item Imai, K., Keele, L., and Yamamoto, T. (2010). Identification, Inference, and Sensitivity Analysis for Causal Mediation Effects.  {\it{Statistical Science}}, 25, 51-71.
\item Kraemer, H., Wilson, G. and Fairburn, C. (2002).  Mediators and moderators of treatment effects in randomized clinical trials.  {\it{Archives of General Psychiatry}}, 59, 877--883.
\item MacKinnon, D.P., Lockwood, C.M., Hoffman, J.M., West, S.G. and Sheets, V. (2002).  A comparison of methods to test mediation and other intervening variable effects.  {\it{Psychological Methods}}, 7, 83-104.
\item Shepherd, B.E., Gilbert, P.B. and Mehrotra, D.V. (2007).  Eliciting a counterfactual sensitivity parameter.  {\it{The American Statistician}}, 61, 56-63.  
\item Stock, J.H., Wright, J.H. and Yogo, M. (2002).  A survey of weak instruments and weak identification in generalized method of moment models. {\it{Journal of Business and Economic Statistics}}, 20, 519-529.
\item Ten Have, T.R., Joffe, M.M., Lynch, K.G., Brown, G.K., Maisto, S.A. and Beck, A.T. (2007).  Causal mediation analyses with rank preserving models.  {\it{Biometrics}}, 63, 926-934.
\item Ten Have, T.R. and Joffe, M.M. (2010).  A review of causal estimation of effects in mediation analyses.  {\it{Biometrics}}, 63, 926-934.
\item Vansteelandt, S. (2009).  Estimating direct effects in cohort and case-control studies.  {\it{Epidemiology}}, 20, 851-860.
\item White, H. (1982).  Instrumental variables regression with independent observations.  {\it{Econometrica}}, 50, 483-499.
\item Wooldridge, J. (2002).  {\it{Econometric Analysis of Cross Section and Panel Data}}.  MIT Press, Cambridge.
\end{description}

\end{document}